\documentclass[twocolumn,pre,showpacs,preprintnumbers,amsmath,amssymb]{revtex4}

\usepackage{graphicx}
\usepackage{dcolumn}
\usepackage{bm}

\begin{document}
\title{\textbf{\textrm{Statistical thermodynamics of a two dimensional relativistic gas}}}

\author{Afshin Montakhab}
    \email{montakhab@shirazu.ac.ir}

\author{Malihe Ghodrat}
\author{Mahmood Barati}

\affiliation{Department of Physics, College of Sciences, Shiraz
    University, Shiraz 71454, Iran}

\date{\today}

\begin{abstract}
In this article we study a fully relativistic model of a two
dimensional hard-disk gas.  This model avoids the general problems
associated with relativistic particle collisions and is therefore
an ideal system to study relativistic effects in statistical
thermodynamics. We study this model using molecular-dynamics
simulation, concentrating on the velocity distribution functions.
We obtain results for $x$ and $y$ components of velocity in the
rest frame ($\Gamma$) as well as the moving frame ($\Gamma'$). Our
results confirm that J\"{u}ttner distribution is the correct
generalization of Maxwell-Boltzmann distribution. We obtain the
same ``temperature'' parameter $\beta$ for both frames consistent
with a recent study of a limited one-dimensional model. We also
address the controversial topic of temperature transformation. We
show that while local thermal equilibrium holds in the moving
frame, relying on statistical methods such as distribution
functions or equipartition theorem are ultimately inconclusive in
deciding on a correct temperature transformation law (if any).
\end{abstract}

\maketitle
\section{Introduction}
The question of how thermodynamic properties transform in a moving
coordinate system were raised soon after Einstein's fundamental
paper in 1905 \cite{Ein05}. In no more than half a century the
introduction of several relativistically consistent generalization
of thermodynamics led to such a confusing atmosphere in which one
could not decide whether a moving body appears cooler, hotter, or
at the same temperature as the body at rest. The most cited view
is presented by Planck \cite{Pla07} and Einstein \cite{Ein07}, who
believed that temperature of a moving body would be Lorentz
contracted. A different view was proposed later by some authors
notably Ott \cite{Ott63} and Arzeli\'{e}s \cite{Arz69}, suggesting
that a body in motion would appear relatively hot. Finally, in
1966 Landsberg \cite{Lan66,Lan67} put forth the third suggestion,
namely, the Lorentz-invariant temperature view. However, 30 years
later Landsberg and Matsas \cite{Lan96,Lan04} and recently Sewell
\cite{Sew08} proposed another view, that of nonexistence of
universal Lorentz transformation of temperature that further
intensified the controversies over the subject.

Since its early days relativistic thermodynamics has changed from
a theoretically interesting problem to a practically important
subject due to its application in the proper interpretation of
experiments in high energy and astrophysics
\cite{Hee06,Ito98,Die06}. Nevertheless, there is still no
consensus on many features of this theory. One reason for the
ongoing discussion is the lack of experimental evidences or
numerical investigations. Among few exceptions is an interesting
paper by Cubero et al. \cite{Cub07} who have shown that a simple
one-dimensional model of relativistic dynamics favors J\"{u}ttner
distribution function \cite{Jut11} as the correct generalization
of Maxwell-Boltzmann (MB) distribution,
\begin{equation}\label{Jut}
f_{J}(\mathbf{v})=m^{d}\gamma(\mathbf{v})^{2+d} \exp[-\beta_{J} m
\gamma(\mathbf{v})]/Z_{J},
\end{equation}
where $d$ is dimension, $Z_{J}$ is normalization constant,
$E=m\gamma(\mathbf{v})$ is relativistic energy, and
$\gamma(\mathbf{v})=(1-v^{2})^{-1/2}$ is the Lorentz factor in
natural units with speed of light $c=1$.

Although the model used in \cite{Cub07} is one-dimensional and
lacks many features of a real physical system, it provides strong
evidence against other generalizations of the Maxwellian,
especially the ``modified'' J\"{u}ttner function
\cite{Hor81,Hor89,Dun07},
\begin{equation}\label{MJut}
f_{MJ}(\mathbf{v})=\frac{m^{d}}{Z_{MJ}}\frac{\gamma(\mathbf{v})^{2+d}}{m
\gamma(\mathbf{v})} \exp[-\beta_{MJ} m \gamma(\mathbf{v})].
\end{equation}

J\"{u}ttner distribution can be used as the cornerstone of our
understanding of relativistic statistical mechanics in the same
manner that MB distribution illuminates the underlying microscopic
roots of classical thermodynamics. The most challenging step,
however, is defining a proper thermometer in order to relate the
Lagrange multiplier $\beta$ to the temperature of the system. This
problem is mostly treated as trivial in the literature but one
should note that, the correct transformation of temperature, like
any other quantity, depends crucially on the practical methods we
implement for its measurement.

Here, we model a two-dimensional (2D) gaseous system with
realistic features which at the same time allows for
implementation of full relativistic dynamics. Since this model is
both realistic and fully relativistic, it can be used as an ideal
numerical laboratory in order to investigate many issues
concerning relativistic generalization of statistical
thermodynamics. Using standard relativistic transformations, we
obtain directional distribution functions for both the rest as
well as the moving frame. We study these functions numerically
using molecular-dynamics simulations of our 2D model. Our results
indicate that J\"{u}ttner distribution is the correct
generalization. We also show that the same temperature parameter
is obtained in both frames. Finally we discuss the implication of
our results for a proper temperature transformation. In this
regard, while verifying the important concept of local thermal
equilibrium, we argue that these methods are ultimately
inconclusive on deciding a correct temperature transformation law.

\section{Model}
We propose to study an idealized two-dimensional system of
impenetrable hard disks with purely repulsive binary interaction
$U(r)$,
\begin{equation}\label{pot}
U(r)=\left\{%
\begin{array}{cc}
  +\infty, & \; r<\sigma \\
  0, & \; r\geq\sigma. \\
\end{array}
\right.
\end{equation}
The disk-like particles move in straight lines at constant speed
and change their momenta instantaneously when they touch at
distance $\sigma$ \cite{ref}. Hence, in order to simulate the
dynamics, we must find the next collision and compute the changes
in momenta of the colliding pair, considering the relativistic
laws of conservation of energy and momentum in two-dimensional
space. In order to solve this problem exactly, we add the
assumption that when two hard disks collide, the force is exerted
along the line connecting their centers,
$\mathbf{r}_{ij}=\mathbf{r}_{i}-\mathbf{r}_{j}$. Therefore, the
components of momenta perpendicular to $\mathbf{r}_{ij}$ remain
unchanged ($\hat{p}_{i,\perp}=p_{\,i,\perp}$) and the parallel
components change in the same way as the one-dimensional case
\cite{JDun07},
\begin{eqnarray}\label{dyn}
    \nonumber \hat{p}_{\,i,\parallel}=\gamma(v_{cm})^{2}[2v_{cm}E_{i}-(1+v_{cm}^{2})p_{\,i,\parallel}],\\
    \hat{E}_{i}=\gamma(v_{cm})^{2}[(1+v_{cm}^{2})E_{i}-2v_{cm}p_{i,\parallel}],
\end{eqnarray}
where hatted quantities refer to momenta after collision and
$v_{cm}=(p\,_{i,\parallel}+p_{j,\parallel})/(E_{i}+E_{j})$ is the
collision invariant, relativistic center-of-mass velocity of the
two particles. With the same rules for particle $j$, a
deterministic, time-reversible canonical transformation at each
collision is defined. The additional assumption means that
particles do not slide on each other when they collide. In
contrast to the one-dimensional model, such elastic binary
collisions lead to equilibrium even if colliding particles carry
the same rest masses \cite{Cub07}. In our simulation we have used
$N$ particles of equal rest masses $m$ that are constrained to
move in a square box of linear size $L$. We use periodic boundary
condition. Note that in order to simulate a stationary system in
the rest frame, the center-of-mass momentum must be put to zero
manually. This condition would automatically be satisfied (if not
at each instant but at least on time average) if fixed reflecting
walls were used \cite{Hai92}.

\begin{figure}

  \includegraphics[width=8cm]{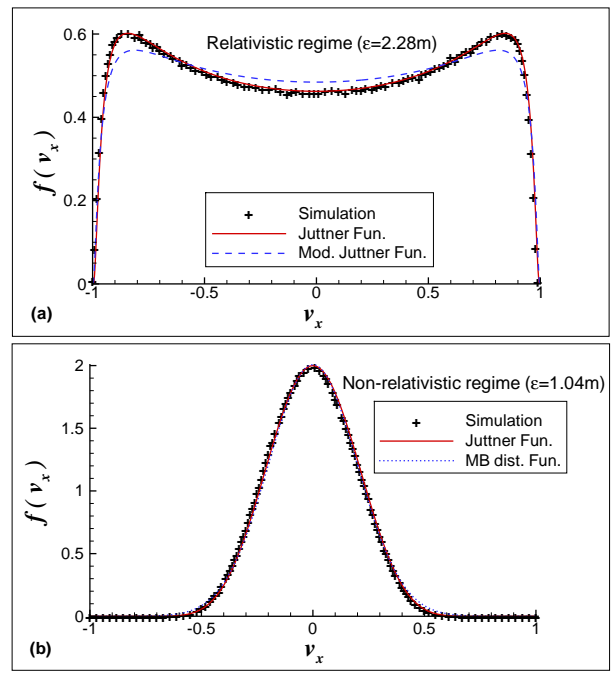}\\
  \caption{Equilibrium velocity distributions in the rest frame $\Gamma$:
  numerically obtained x component of single-particle velocity
  distribution(+) from simulation of $N=100$ particles of mass $m=0.1$.
  (a) Here, $\epsilon=2.28m$ and the corresponding temperature parameters are
  $\beta_{J}=11.4$, $\beta_{MJ}=7.8$. (b) $\epsilon=1.04m$,
  $\beta_{J}=253.1$ and $\beta_{MB}=259.8$.
  A significant deviation from modified J\"{u}ttner function is evident in the relativistic regime.
  Similar results are obtained for $f(v_{y})$.}
\end{figure}

\section{Results}
\subsection{Rest frame}
In order to obtain the equilibrium state of the system we let the
two-dimensional gas equilibrate (typically after $10^{2}N$
collisions) and measure velocities of particles at equal times
with respect to laboratory frame. To collect more data, we
repeated this procedure every $10N$ collisions. Simulation results
for $N=100$ particles are presented. Particles are initially
placed on a square lattice of constant $L/\sqrt{N}$ and velocities
are chosen randomly in $\delta$-vicinity (with $\delta$ a small
number) of $|\mathbf{v_{0}}|=\sqrt{1-(1/\eta)^{2}}$, corresponding
to mean energy per particle $\epsilon=\eta m$. As $\eta\rightarrow
1$, the kinetic energy becomes small compared to the rest mass
energy, recovering the dynamics of (classical) non-relativistic
model. The $\eta\rightarrow \infty$ limit, on the other hand,
corresponds to a highly relativistic model.

Theoretically, the $x$($y$) component of J\"{u}ttner and modified
J\"{u}ttner velocity distribution is obtained by integrating
Eqs.(\ref{Jut}) and (\ref{MJut}) over $v_{y}$($v_{x}$),
\begin{equation}\label{Jdis}
  f_{J}(v_{x})=\frac{m^{2}}{Z_{J}}\gamma(v_{x})^{3}[K_{2}(\beta_{J} m \gamma(v_{x}))+
  K_{0}(\beta_{J} m \gamma(v_{x}))],
\end{equation}
\begin{equation}\label{MJdis}
    f_{MJ}(v_{x})=\frac{2m}{Z_{MJ}}\gamma(v_{x})^{2}K_{1}(\beta_{MJ} m
    \gamma(v_{x})),
\end{equation}
with $K_{n}$ denoting modified Bessel functions of the second kind
\cite{Abr72}. Here, the parameter $\beta_{J/MJ}$ is determined by
means of the following procedure: we have, for the average energy,
\begin{equation}\label{ene}
    \epsilon=E_{tot}/N=\int_{|\mathbf{v}|<1}d^{d}\mathbf{v}f(\mathbf{v})m\gamma(\mathbf{v}).
\end{equation}
Computing the right hand side (rhs) of Eq.(\ref{ene}) for
J\"{u}ttner and modified J\"{u}ttner in the two-dimensional case
gives $\text{rhs}_{J}=(\beta^{2}m^{2}+2\beta m+2)/(\beta(\beta m
+1))$ and $\text{rhs}_{MJ}=1/\beta +m$, respectively. By inserting
$E_{tot}$, $N$ and $m$ into these equations, the parameter
$\beta_{J/MJ}$ consistent with J\"{u}ttner and modified
J\"{u}ttner velocity distribution is uniquely determined.

We now check these results by considering two cases with
$\eta=1.04$ and $\eta=2.28$. As shown in Fig.1, the obtained
single particle distribution of velocity $x$-component (+) agrees
with J\"{u}ttner function (solid lines) in both regimes and
converges to MB distribution in the non-relativistic limit. A
significant deviation from modified J\"{u}ttner distribution
(dashed lines) is also evident. Exact same diagrams are obtained
for $y$-component of velocity (not shown). In particular, the
$y$-component results were fitted with the same parameter $\beta$
as the $x$-component data. This shows that our system has
equilibrated properly through successive collisions.
\begin{figure}
  \includegraphics[width=8cm]{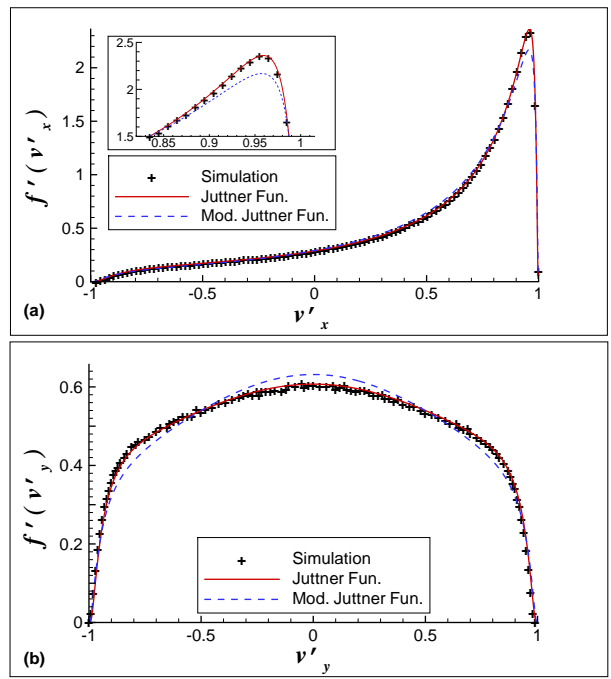}\\
  \caption{Equilibrium velocity distributions in the moving frame
  $\Gamma'$ with relative velocity $u=0.5$. The system parameters
  are the same as Fig.1(a) in particular $\beta_{J}=11.4$ and
  $\beta_{MJ}=7.8$. Note the breaking of symmetry in part
  (a), where the inset shows more details of the peak.}
\end{figure}

\smallskip
\subsection{Moving frame}
We now turn to the more interesting question of equilibrium
velocity distribution of a relativistic gas in motion. For this,
we examine the system from the point of view of an observer who
sees that the system's frame, $\Gamma'$, is moving with a uniform
velocity $u$ in $x$-direction with respect to his rest frame,
$\Gamma$. Using the entropy maximization principle, the
single-particle distribution will be determined by an additional
constraint on the system, namely, that of a definite total
momentum $\mathbf{p}'$ \cite{pat72},

\begin{equation}\label{movfv}
    f'_{J}(\mathbf{v}')=\frac{m^{d}\gamma(\mathbf{v}')^{d+2}}{\gamma(u)Z_{J}}\exp[\beta_{J}
\gamma(u)m\gamma(\mathbf{v}')(1-\mathbf{u}.\mathbf{v}')]
\end{equation}

\begin{equation}\label{movfmv}
    f'_{MJ}(\mathbf{v}')=\frac{m^{d}\gamma(\mathbf{v}')^{d+2}}{\gamma(u)Z_{MJ}}
    \frac{\exp[\beta_{MJ}\gamma(u)m\gamma(\mathbf{v}')(1-\mathbf{u}.\mathbf{v}')]}
    {\gamma(u)m\gamma(\mathbf{v}')(1-\mathbf{u}.\mathbf{v}')}
\end{equation}
The primed quantities are measured in the moving frame and the
additional $\gamma(u)$ term in denominator is due to the
contraction of the moving box that encloses the system
\cite{Kam69}. Figure 2 shows the results for a system similar to
Fig.1(a) with $u=0.5$. Note that here, $x(y)$ component of
velocities are measured $\Gamma'$-simultaneously. The solid and
dashed lines are the velocity $x(y)$ component of velocity
distribution obtained, respectively, by integrating
Eqs.(\ref{movfv}) and (\ref{movfmv}) over $v_{y}(v_{x})$,e.g.,
\begin{eqnarray}\label{Jdist}
   \nonumber f'_{J}(v'_{x})=\frac{m^{2}}{Z_{J}}\gamma(v'_{x})^{3}[K_{2}(\beta m \gamma(u)\gamma(v'_{x})(1-uv'_{x}))+\\
  K_{0}(\beta m \gamma(u)\gamma(v'_{x})(1-uv'_{x}))],
\end{eqnarray}

Other components of these distributions cannot be obtained in
closed form and are therefore plotted in Fig.2 using numerical
integration. The parameter $\beta$ used to fit data in Fig.2 is
the same as that of the rest frame. Note that this parameter is
obtained exactly as a function of system parameters $(m,\epsilon)$
in $\Gamma$. However, in $\Gamma'$, $\beta$ is a fitting parameter
which turns out to be the same as that in $\Gamma$.

Therefore as clearly seen from numerical results, our two
dimensional model shows J\"{u}ttner distribution as the correct
relativistic version of MB distribution. The fact that the same
parameter $\beta$ is obtained from both $x$ and $y$ component
velocities shows equilibration. However, more importantly, the
fact that same $\beta$ is obtained from both $\Gamma$ and
$\Gamma'$ frames seems to indicate the invariance of temperature
consistent with earlier work of Landsberg \cite{Lan66} and
previous simulation results \cite{Cub07}. We now discuss if this
agreement can shed light on the long-lasting question of how
temperature transforms in a moving frame.

\smallskip
\section{Proper Thermometer}
A commonly used definition of equilibrium temperature in the literature is
$T=(k_{B}\beta)^{-1}$, where $\beta$ is the Lagrange multiplier
emerging in the velocity distribution function. One may use this
definition and the equality of parameter $\beta$ in moving and
rest frame (Fig.2) to deduce that temperature is Lorentz
invariant, $T'=T$ \cite{Cub07}. However, there is another point of
view which is also consistent with our results. Comparing
Eq.(\ref{movfv}) with the general form of the distribution
function as $f\propto
e^{-(\alpha+\beta\varepsilon+\omega.\mathbf{P})}$, one is led to
believe that the Lagrange multiplier is $\gamma(u)\beta$ which
gives $T'=T/\gamma(u)$ \cite{Pla07,Ein07,pat72}.

To see this point better, note that the above definition of
temperature has its roots in the equipartition theorem as well as
basic thermodynamic relation, $dE=TdS-PdV+\mu dN$. The
relativistic version of these methods is widely used in order to
find the temperature transformation law
\cite{Pla07,Pau58,pat72,Ott63,Arz69,Cub07}. However, despite the
popularity of these approaches, they are not decisive either
\cite{Kam68,Yue70,Cal71}. To illuminate, consider the
Lorentz-invariant equipartition theorem for a system moving with
velocity $u$ parallel to $x$ axis \cite{Lan67},
\begin{equation}\label{eupar}
    \ll \frac{p'^{2}_{ix}}{m'_{i}}-up'_{ix}\gg =
    \ll \frac{p'^{2}_{iy}}{m'_{i}}\gg = \frac{1}{\beta\gamma(u)}
\end{equation}
where the primed quantities are measured in the moving frame
[e.g., $m'_{i}=\gamma(\mathbf{v}_{i})m_{i}$] and averages
$\ll\ldots\gg$ are taken $\Gamma'$-simultaneously. One may apply
either hypothesis that $\ll p'^{2}_{ix}/m'_{i}-up'_{ix}\gg$ or
$\ll\gamma(u)(p'^{2}_{ix}/m'_{i}-up'_{ix})\gg $ are the
statistical thermometer of the moving system and find it
compatible with the generalized theorem. Thus, as Landsberg has
mentioned in \cite{Lan67}: ``the argument from equipartition does
not enable one to discriminate on theoretical grounds between
accepted theory and Lorentz-invariant temperature.''

\begin{figure}
  \includegraphics[width=8cm]{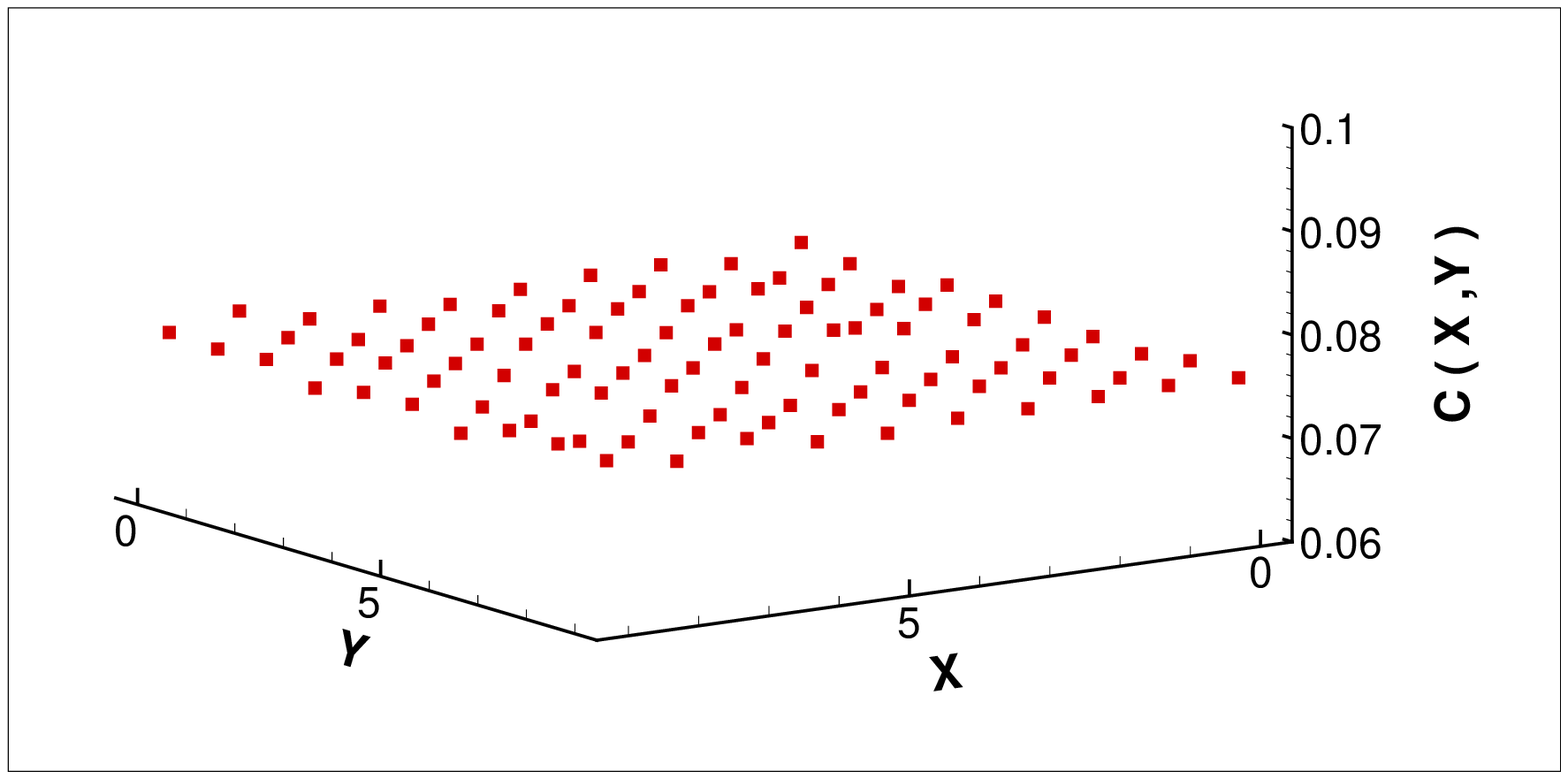}\\
  \caption{Local thermal equilibrium in the box as seen by a moving observer: The profile
  $\mathcal{C}(X,Y)$ is measured $\Gamma'$-simultaneously for the same system as in Fig.2
  with $L_{x}$ and $L_{y}$, each divided into ten parts. The expected value of $[\beta\gamma(u)]^{-1}$
  is 0.076 which agrees well with the obtained value of $\mathcal{C}$ throughout the lattice.}
\end{figure}

Furthermore, in some recent papers it is claimed that there exists
no universal and continuous Lorentz transformation of temperature
at all \cite{Lan96,Lan04}. The argument is based on the fact that
black body radiation of a moving body is direction dependent.
Therefore, a bath which is thermal in an inertial frame is
non-thermal in a moving frame. Does local thermal equilibrium
(LTE) hold in our system? To check, we divide the box into $n$
cells of area $\Delta A'=(L'_{x}L'_{y})/n^{2}$ and calculate the
quantity $\mathcal{C}=\ll p'^{2}_{ix}/m'_{i}-up'_{ix}\gg$, which
we consider to be proportional to temperature, in each cell. The
numerical result shown in Fig.3, indicates that each cell, as seen
by a moving observer, is characterized by a constant value which
coincides with the value of $[\beta\gamma(u)]^{-1}$. This
indicates that LTE, which is the necessary condition to introduce
a well defined temperature, is fulfilled at least for our model.

\section{Concluding Remarks}
It seems that the longstanding issue of relativistic
thermodynamics is related to the longstanding issue of
irreversibility in thermodynamics. The lack of consensus on these
issues is related to the lack of concise mapping between dynamical
description of a system on one hand and a thermodynamic
description on the other. We cannot define temperature (or
entropy) as an exact function of dynamical variables. In this work
we have modeled a useful, realistic system of a relativistic gas
which overcomes the difficulties associated with implementation of
particle interactions in a relativistically consistent manner
\cite{Cub07, Whe49}. We have shown that J\"{u}ttner function is
the correct velocity distribution function in both rest and moving
frames, with components either along or perpendicular to the
relative velocity $u$. Furthermore, our results indicate that,
with a certain definition of statistical thermometer, one can
choose $\beta'=\beta$, i.e., a Lorentz-invariant temperature,
without running into inconsistencies. However,
$\beta'=\gamma(u)\beta$ could just as well be argued to be a valid
choice, depending on a choice of thermometer. Such
inconclusiveness inherit in statistical analysis like ours leads
one to consider a covariant formulation of thermodynamics where
temperature is generalized to a tensorial quantity whose
transformation is no longer an issue \cite{Kam68,Isr86}. In this
view thermodynamic temperature is considered as a \emph{proper}
feature of a thermodynamic system, much like mass in relativistic
mechanics \cite{Tay92}.

\begin{acknowledgments}
The authors kindly acknowledge the support of Shiraz University
Research Council. We would also like to acknowledge the criticism
of an anonymous referee which helped clarify some points regarding
non-existence of temperature transformation.
\end{acknowledgments}

\bibliographystyle{apsrev}
\bibliography{xbib}

\begin{thebibliography}{31}
\expandafter\ifx\csname natexlab\endcsname\relax\def\natexlab#1{#1}\fi
\expandafter\ifx\csname bibnamefont\endcsname\relax
  \def\bibnamefont#1{#1}\fi
\expandafter\ifx\csname bibfnamefont\endcsname\relax
  \def\bibfnamefont#1{#1}\fi
\expandafter\ifx\csname citenamefont\endcsname\relax
  \def\citenamefont#1{#1}\fi
\expandafter\ifx\csname url\endcsname\relax
  \def\url#1{\texttt{#1}}\fi
\expandafter\ifx\csname urlprefix\endcsname\relax\def\urlprefix{URL }\fi
\providecommand{\bibinfo}[2]{#2}
\providecommand{\eprint}[2][]{\url{#2}}

\bibitem[{\citenamefont{Einstein}(1905)}]{Ein05}
\bibinfo{author}{\bibfnamefont{A.}~\bibnamefont{Einstein}},
  \bibinfo{journal}{Ann. Phys.} \textbf{\bibinfo{volume}{322}},
  \bibinfo{pages}{891} (\bibinfo{year}{1905}).

\bibitem[{\citenamefont{Planck}(1908)}]{Pla07}
\bibinfo{author}{\bibfnamefont{M.}~\bibnamefont{Planck}},
  \bibinfo{journal}{Ann. Phys.} \textbf{\bibinfo{volume}{331}},
  \bibinfo{pages}{1} (\bibinfo{year}{1908}).

\bibitem[{\citenamefont{Einstein}(1907)}]{Ein07}
\bibinfo{author}{\bibfnamefont{A.}~\bibnamefont{Einstein}},
  \bibinfo{journal}{Jahrb. Radioakt. Elektron.} \textbf{\bibinfo{volume}{4}},
  \bibinfo{pages}{411} (\bibinfo{year}{1907}).

\bibitem[{\citenamefont{Ott}(1963)}]{Ott63}
\bibinfo{author}{\bibfnamefont{H.}~\bibnamefont{Ott}}, \bibinfo{journal}{Z.
  Phys.} \textbf{\bibinfo{volume}{175}}, \bibinfo{pages}{70}
  (\bibinfo{year}{1963}).

\bibitem[{\citenamefont{Arzelies}(1969)}]{Arz69}
\bibinfo{author}{\bibfnamefont{H.}~\bibnamefont{Arzelies}},
  \emph{\bibinfo{title}{Thermodynamique Relativiste et Quantique}}
  (\bibinfo{publisher}{Gauthier-villars}, \bibinfo{address}{Paris},
  \bibinfo{year}{1969}).

\bibitem[{\citenamefont{Landsberg}(1966)}]{Lan66}
\bibinfo{author}{\bibfnamefont{P.~T.} \bibnamefont{Landsberg}},
  \bibinfo{journal}{Nature (London)} \textbf{\bibinfo{volume}{212}},
  \bibinfo{pages}{571} (\bibinfo{year}{1966}).

\bibitem[{\citenamefont{Landsberg}(1967)}]{Lan67}
\bibinfo{author}{\bibfnamefont{P.~T.} \bibnamefont{Landsberg}},
  \bibinfo{journal}{Nature (London)} \textbf{\bibinfo{volume}{214}},
  \bibinfo{pages}{903} (\bibinfo{year}{1967}).

\bibitem[{\citenamefont{Landsberg and Matsas}(1996)}]{Lan96}
\bibinfo{author}{\bibfnamefont{P.~T.} \bibnamefont{Landsberg}}
  \bibnamefont{and} \bibinfo{author}{\bibfnamefont{G.~E.~A.}
  \bibnamefont{Matsas}}, \bibinfo{journal}{Phys. Lett. A}
  \textbf{\bibinfo{volume}{223}}, \bibinfo{pages}{401} (\bibinfo{year}{1996}).

\bibitem[{\citenamefont{Landsberg and Matsas}(2004)}]{Lan04}
\bibinfo{author}{\bibfnamefont{P.~T.} \bibnamefont{Landsberg}}
  \bibnamefont{and} \bibinfo{author}{\bibfnamefont{G.~E.~A.}
  \bibnamefont{Matsas}}, \bibinfo{journal}{Physica A}
  \textbf{\bibinfo{volume}{340}}, \bibinfo{pages}{92} (\bibinfo{year}{2004}).

\bibitem[{\citenamefont{Sewell}(2008)}]{Sew08}
\bibinfo{author}{\bibfnamefont{G.~L.} \bibnamefont{Sewell}},
  \bibinfo{journal}{J. Phys. A: Math Theor.} \textbf{\bibinfo{volume}{41}},
  \bibinfo{pages}{382003} (\bibinfo{year}{2008}).

\bibitem[{\citenamefont{van Hees et~al.}(2006)\citenamefont{van Hees, Greco,
  and Rapp}}]{Hee06}
\bibinfo{author}{\bibfnamefont{H.}~\bibnamefont{van Hees}},
  \bibinfo{author}{\bibfnamefont{V.}~\bibnamefont{Greco}}, \bibnamefont{and}
  \bibinfo{author}{\bibfnamefont{R.}~\bibnamefont{Rapp}},
  \bibinfo{journal}{Phys. Rev. C} \textbf{\bibinfo{volume}{73}},
  \bibinfo{pages}{034913} (\bibinfo{year}{2006}).

\bibitem[{\citenamefont{Itoh et~al.}(1998)\citenamefont{Itoh, Kohyama, and
  Nozawa}}]{Ito98}
\bibinfo{author}{\bibfnamefont{N.}~\bibnamefont{Itoh}},
  \bibinfo{author}{\bibfnamefont{Y.}~\bibnamefont{Kohyama}}, \bibnamefont{and}
  \bibinfo{author}{\bibfnamefont{S.}~\bibnamefont{Nozawa}},
  \bibinfo{journal}{Astrophys. J.} \textbf{\bibinfo{volume}{502}},
  \bibinfo{pages}{7} (\bibinfo{year}{1998}).

\bibitem[{\citenamefont{Dieckmann et~al.}(2006)\citenamefont{Dieckmann, Drury,
  and Shukla}}]{Die06}
\bibinfo{author}{\bibfnamefont{M.~E.} \bibnamefont{Dieckmann}},
  \bibinfo{author}{\bibfnamefont{L.}~\bibnamefont{Drury}}, \bibnamefont{and}
  \bibinfo{author}{\bibfnamefont{P.~K.} \bibnamefont{Shukla}},
  \bibinfo{journal}{New J. Phys.} \textbf{\bibinfo{volume}{8}},
  \bibinfo{pages}{40} (\bibinfo{year}{2006}).

\bibitem[{\citenamefont{Cubero et~al.}(2007)\citenamefont{Cubero,
  Casado-Pascual, Dunkel, Talkner, and H\"{a}nggi}}]{Cub07}
\bibinfo{author}{\bibfnamefont{D.}~\bibnamefont{Cubero}},
  \bibinfo{author}{\bibfnamefont{J.}~\bibnamefont{Casado-Pascual}},
  \bibinfo{author}{\bibfnamefont{J.}~\bibnamefont{Dunkel}},
  \bibinfo{author}{\bibfnamefont{P.}~\bibnamefont{Talkner}}, \bibnamefont{and}
  \bibinfo{author}{\bibfnamefont{P.}~\bibnamefont{H\"{a}nggi}},
  \bibinfo{journal}{Phys. Rev. Lett.} \textbf{\bibinfo{volume}{99}},
  \bibinfo{pages}{170601} (\bibinfo{year}{2007}).

\bibitem[{\citenamefont{J\"{u}ttner}(1911)}]{Jut11}
\bibinfo{author}{\bibfnamefont{F.}~\bibnamefont{J\"{u}ttner}},
  \bibinfo{journal}{Ann. Phys.} \textbf{\bibinfo{volume}{34}},
  \bibinfo{pages}{856} (\bibinfo{year}{1911}).

\bibitem[{\citenamefont{Horwitz et~al.}(1981)\citenamefont{Horwitz, Schieve,
  and Piron}}]{Hor81}
\bibinfo{author}{\bibfnamefont{L.~P.} \bibnamefont{Horwitz}},
  \bibinfo{author}{\bibfnamefont{W.~C.} \bibnamefont{Schieve}},
  \bibnamefont{and} \bibinfo{author}{\bibfnamefont{C.}~\bibnamefont{Piron}},
  \bibinfo{journal}{Ann. Phys. (N.Y.)} \textbf{\bibinfo{volume}{137}},
  \bibinfo{pages}{306} (\bibinfo{year}{1981}).

\bibitem[{\citenamefont{Horwitz et~al.}(1989)\citenamefont{Horwitz, Shashoua,
  and Schieve}}]{Hor89}
\bibinfo{author}{\bibfnamefont{L.~P.} \bibnamefont{Horwitz}},
  \bibinfo{author}{\bibfnamefont{S.}~\bibnamefont{Shashoua}}, \bibnamefont{and}
  \bibinfo{author}{\bibfnamefont{W.~C.} \bibnamefont{Schieve}},
  \bibinfo{journal}{physica A} \textbf{\bibinfo{volume}{161}},
  \bibinfo{pages}{300} (\bibinfo{year}{1989}).

\bibitem[{\citenamefont{Dunkel et~al.}(2007)\citenamefont{Dunkel, Talkner, and
  H\"{a}nggi}}]{Dun07}
\bibinfo{author}{\bibfnamefont{J.}~\bibnamefont{Dunkel}},
  \bibinfo{author}{\bibfnamefont{P.}~\bibnamefont{Talkner}}, \bibnamefont{and}
  \bibinfo{author}{\bibfnamefont{P.}~\bibnamefont{H\"{a}nggi}},
  \bibinfo{journal}{New J. Phys.} \textbf{\bibinfo{volume}{9}},
  \bibinfo{pages}{144} (\bibinfo{year}{2007}).

\bibitem[{ref()}]{ref}
\bibinfo{note}{A less frame-dependent form of Eq.(3) would define such a
  potential in the center-of-mass which however would not change our results in
  any significant way.}

\bibitem[{\citenamefont{Dunkel and H\"{a}nggi}(2007)}]{JDun07}
\bibinfo{author}{\bibfnamefont{J.}~\bibnamefont{Dunkel}} \bibnamefont{and}
  \bibinfo{author}{\bibfnamefont{P.}~\bibnamefont{H\"{a}nggi}},
  \bibinfo{journal}{Physica A} \textbf{\bibinfo{volume}{374}},
  \bibinfo{pages}{559} (\bibinfo{year}{2007}).

\bibitem[{\citenamefont{Haile}(1992)}]{Hai92}
\bibinfo{author}{\bibfnamefont{J.~M.} \bibnamefont{Haile}},
  \emph{\bibinfo{title}{Molecular Dynamics Simulation}}
  (\bibinfo{publisher}{Wiley}, \bibinfo{year}{1992}), \bibinfo{note}{see
  Appendix B.}

\bibitem[{\citenamefont{Abramowitz and Stegun}(1972)}]{Abr72}
\bibinfo{editor}{\bibfnamefont{M.}~\bibnamefont{Abramowitz}} \bibnamefont{and}
  \bibinfo{editor}{\bibfnamefont{I.~A.} \bibnamefont{Stegun}}, eds.,
  \emph{\bibinfo{title}{Handbook of Mathematical Functions}}
  (\bibinfo{publisher}{Dover}, \bibinfo{address}{New York},
  \bibinfo{year}{1972}).

\bibitem[{\citenamefont{Pathria}(1972)}]{pat72}
\bibinfo{author}{\bibfnamefont{R.~K.} \bibnamefont{Pathria}},
  \emph{\bibinfo{title}{Statistical Mechanics}} (\bibinfo{publisher}{Pergamon
  Press}, \bibinfo{address}{Oxford}, \bibinfo{year}{1972}),
  \bibinfo{edition}{1st} ed.

\bibitem[{\citenamefont{van Kampen}(1969)}]{Kam69}
\bibinfo{author}{\bibfnamefont{N.~G.} \bibnamefont{van Kampen}},
  \bibinfo{journal}{Physica (Utrecht)} \textbf{\bibinfo{volume}{43}},
  \bibinfo{pages}{244} (\bibinfo{year}{1969}).

\bibitem[{\citenamefont{Pauli}(1958)}]{Pau58}
\bibinfo{author}{\bibfnamefont{W.}~\bibnamefont{Pauli}},
  \emph{\bibinfo{title}{Theory of Relativity}} (\bibinfo{publisher}{Pergamon
  Press}, \bibinfo{address}{London}, \bibinfo{year}{1958}).

\bibitem[{\citenamefont{van Kampen}(1968)}]{Kam68}
\bibinfo{author}{\bibfnamefont{N.~G.} \bibnamefont{van Kampen}},
  \bibinfo{journal}{Phys. Rev.} \textbf{\bibinfo{volume}{173}},
  \bibinfo{pages}{295} (\bibinfo{year}{1968}).

\bibitem[{\citenamefont{Yuen}(1970)}]{Yue70}
\bibinfo{author}{\bibfnamefont{C.~K.} \bibnamefont{Yuen}},
  \bibinfo{journal}{Am. J. Phys.} \textbf{\bibinfo{volume}{38}},
  \bibinfo{pages}{246} (\bibinfo{year}{1970}).

\bibitem[{\citenamefont{Callen and Horwitz}(1971)}]{Cal71}
\bibinfo{author}{\bibfnamefont{H.}~\bibnamefont{Callen}} \bibnamefont{and}
  \bibinfo{author}{\bibfnamefont{G.}~\bibnamefont{Horwitz}},
  \bibinfo{journal}{Am. J. Phys.} \textbf{\bibinfo{volume}{39}},
  \bibinfo{pages}{938} (\bibinfo{year}{1971}).

\bibitem[{\citenamefont{Wheeler and Feynmann}(1949)}]{Whe49}
\bibinfo{author}{\bibfnamefont{J.~A.} \bibnamefont{Wheeler}} \bibnamefont{and}
  \bibinfo{author}{\bibfnamefont{R.~P.} \bibnamefont{Feynmann}},
  \bibinfo{journal}{Rev. Mod. Phys.} \textbf{\bibinfo{volume}{21}},
  \bibinfo{pages}{425} (\bibinfo{year}{1949}).

\bibitem[{\citenamefont{Israel}(1986)}]{Isr86}
\bibinfo{author}{\bibfnamefont{W.}~\bibnamefont{Israel}}, \bibinfo{journal}{J.
  Non-Equilb. Thermodyn.} \textbf{\bibinfo{volume}{11}}, \bibinfo{pages}{295}
  (\bibinfo{year}{1986}).

\bibitem[{\citenamefont{Taylor and Wheeler}(1992)}]{Tay92}
\bibinfo{author}{\bibfnamefont{E.~F.} \bibnamefont{Taylor}} \bibnamefont{and}
  \bibinfo{author}{\bibfnamefont{J.~A.} \bibnamefont{Wheeler}},
  \emph{\bibinfo{title}{Spacetime Physics}} (\bibinfo{publisher}{Freemann
  Press}, \bibinfo{address}{San Francisco}, \bibinfo{year}{1992}),
  \bibinfo{edition}{2nd} ed.

\end{thebibliography}
\end{document}